\documentclass[final,3p,times]{elsarticle}

\usepackage{amssymb}
\usepackage[T1]{fontenc}
\usepackage{lmodern}
\usepackage[utf8]{inputenc}
\usepackage{amsmath}
\usepackage{subfigure}

\journal{Journal of Subatomic Particles and Cosmology}

\begin{document}

\begin{frontmatter}



\title{Thermodynamically Consistent Merging of Multidimensional QCD Equations of State}

\author[UH]{Prachi Garella\fnref{presenter}}
\fntext[presenter]{Presenter at SQM 2026.}
\author[UIUC]{Yumu Yang}
\author[UH]{Musa R. Khan}
\author[UH]{Tulio E. Restrepo}
\author[UH,KSU]{Joaquin Grefa}
\author[UH]{Johannes Jahan}
\author[CBPF]{Mauricio Hippert}
\author[UIUC]{Jorge Noronha}
\author[UH]{Claudia Ratti}
\author[RR]{Romulo Rougemont}

\affiliation[UH]{
    organization={Department of Physics, University of Houston},
    city={Houston},
    state={TX},
    postcode={77204},
    country={USA}
}

\affiliation[UIUC]{
    organization={Illinois Center for Advanced Studies of the Universe,
    Department of Physics, University of Illinois Urbana-Champaign},
    city={Urbana},
    state={IL},
    postcode={61801},
    country={USA}
}

\affiliation[KSU]{
    organization={Center for Nuclear Research,
    Department of Physics, Kent State University},
    city={Kent},
    state={OH},
    postcode={44242},
    country={USA}
}

\affiliation[CBPF]{
    organization={Centro Brasileiro de Pesquisas F\'{\i}sicas},
    addressline={Rua Dr.\ Xavier Sigaud 150},
    city={Rio de Janeiro},
    state={RJ},
    postcode={22290-180},
    country={Brazil}
}
\affiliation[RR]{
    organization={Instituto de F\'{\i}sica, Universidade Federal de Goi\'{a}s},
    addressline={Avenida Esperan\c{c}a -- Campus Samambaia},
    city={Goi\^{a}nia},
    state={Goi\'{a}s},
    postcode={74690-900},
    country={Brazil}
}
\begin{abstract}
We present a thermodynamically consistent framework for merging complementary
models into a multidimensional QCD equation of state. An internal mixing
variable is determined by minimizing a single grand potential at fixed
temperature and baryon chemical potential, ensuring thermodynamic consistency
and stability. Interactions between the components allow for a crossover, a
critical endpoint, and a first-order transition. As a proof of principle, we
merge a quantum van der Waals hadron-resonance-gas model with a holographic
Einstein--Maxwell--Dilaton model. The resulting equation of state reproduces
the appropriate description in each regime, agrees well with available
lattice-QCD results, and is suitable for heavy-ion phenomenology over a broad
range of temperature and baryon chemical potential.
\end{abstract}







\end{frontmatter}



\section{Introduction}
\label{Intro}

The equation of state (EoS) of strongly interacting matter is essential for
describing relativistic heavy-ion collisions and dense astrophysical
systems~\cite{MUSES:2023hyz}. These applications require thermodynamic
information over a broad range of temperature $T$ and baryon chemical
potential $\mu_B$, where no single theoretical framework is quantitatively
reliable. Lattice QCD provides first-principles results at $\mu_B=0$ and can
be extended to moderate $\mu_B/T$, but calculations at large baryon density
remain limited by the sign problem~\cite{Bazavov:2014pvz,Borsanyi:2021sxv}.

At low temperature, interacting hadron-resonance-gas models describe the
confined phase but not the deconfined quark--gluon
plasma~\cite{Vovchenko:2016rkn}. Conversely, holographic
Einstein--Maxwell--Dilaton models can reproduce lattice-QCD thermodynamics in
the strongly coupled deconfined regime but do not contain the hadronic degrees
of freedom required at low temperature~\cite{Hippert:2023bel}. A global EoS
must therefore connect these complementary descriptions without spoiling their
thermodynamic structure.

A common approach defines the pressure as a weighted average of two input
equations of state using a prescribed switching
function~\cite{Parotto:2018pwx}. However, derivatives of the
switching function generate additional contributions to densities and
susceptibilities that may produce artificial structures or compromise
thermodynamic stability.

Here, we summarize a thermodynamically consistent framework in which the
relative fraction of the two descriptions is promoted to an internal,
order-parameter-like variable~\cite{Yang:2026brr}. Its equilibrium value is
obtained by minimizing a merged grand-potential density at fixed $T$ and
$\mu_B$, ensuring that all observables follow from a single thermodynamic
potential. The construction can describe a crossover, a critical endpoint,
and a first-order transition. As a proof of principle we apply it to merge quantum
van der Waals hadron-resonance-gas (QvdW-HRG) and holographic
EMD equations of state into a unified description of hadronic and deconfined
matter.

\section{Statistical-mixture framework}
\label{sec:statistical_mixture}

\subsection{Equilibrium construction}

The construction replaces a prescribed switching function by an internal
variable $p\in[0,1]$, representing the relative fractions of two fluids with
pressures $P_1(T,\mu_B)$ and $P_2(T,\mu_B)$~\cite{Yang:2026brr}. In the grand
canonical ensemble, equilibrium requires a stable minimum of the
grand-potential density,
\begin{equation}
    \left.
    \frac{\partial\omega}{\partial p}
    \right|_{T,\mu_B}=0,
    \qquad
    \left.
    \frac{\partial^2\omega}{\partial p^2}
    \right|_{T,\mu_B}\geq0.
    \label{eq:equilibrium_conditions}
\end{equation}
The first condition removes contributions proportional to
$\partial P/\partial p$ from first thermodynamic derivatives, while the second
ensures local stability.

Because a linear combination of the input grand potentials cannot have an
interior minimum for $P_1\neq P_2$, we adopt a statistical-mixture form
motivated by Flory--Huggins solution theory~\cite{flory1953principles},
\begin{equation}
    \omega(T,\mu_B;p)
    =-pP_1-(1-p)P_2
    +a\left[p\ln p+(1-p)\ln(1-p)\right]
    +b\,p(1-p).
    \label{eq:merged_grand_potential}
\end{equation}
The term proportional to $a$ is the entropy of mixing and favors intermediate
values of $p$, whereas the interaction term proportional to $b$ can generate
two competing minima. Denoting the global minimum by
$\bar p(T,\mu_B)$, the equilibrium pressure is
\begin{equation}
    P(T,\mu_B)
    =-\omega\left(T,\mu_B;\bar p(T,\mu_B)\right).
    \label{eq:merged_pressure}
\end{equation}

\subsection{Phase structure and thermodynamics}
\label{subsec:phase_structure}

When $P_1=P_2$, the grand potential is symmetric under
$p\rightarrow1-p$, and $\bar p=1/2$ is a stationary solution. The phase
structure is controlled by the relative values of $a$ and $b$: $b<2a$
produces a crossover, $b=2a$ defines the critical point, and $b>2a$ yields
two degenerate minima and a first-order transition. Motivated by the entropy-of-mixing interpretation, we choose, $a=\frac{T}{\Delta V},\;b=\frac{2T_c}{\Delta V}$
where $\Delta V$ is a phenomenological volume parameter and $T_c$ sets the
critical temperature. If the input pressures become equal at $T=T_t$, the transition is a crossover for $T_t>T_c$ and first order for $T_t<T_c$.

All thermodynamic observables follow from the equilibrium pressure in
Eq.~\eqref{eq:merged_pressure}. Second derivatives receive additional
contributions from fluctuations of $\bar p$; these vanish in the pure limits
$\bar p=0,1$ and diverge at $T=T_c$ and $\bar p=1/2$.
\section{Proof of principle}
\label{sec:proof_of_principle}

\subsection{Input equations of state}
\label{subsec:input_eos}

As a proof of principle, we merge complementary descriptions of the hadronic
and deconfined regimes. The low-temperature sector is described by a QvdW-HRG equation of
state~\cite{Vovchenko:2016rkn}. It includes attractive and repulsive
interactions between baryons and excluded-volume repulsion between mesons.
The baryonic parameters are fixed by nuclear-matter saturation properties,
allowing the model to reproduce the nuclear liquid--gas transition and improve
baryon-number susceptibilities at $\mu_B=0$~\cite{Yang:2026brr}.

The deconfined sector is described by a five-dimensional
Einstein--Maxwell--Dilaton model based on the gauge/gravity
correspondence~\cite{Maldacena:1997re,Hippert:2023bel}. The strongly coupled
plasma is represented by a charged asymptotically AdS$_5$ black brane, with
the dilaton and bulk $U(1)$ field controlling the nonconformal and baryonic
sectors. A Bayesian calibration to lattice-QCD results at $\mu_B=0$ yields
quantitative agreement up to $\mu_B/T=3.5$ and predicts a first-order
transition line ending at a critical point. These two equations of state
therefore provide complementary inputs for testing the merging procedure.

\subsection{Merged equation of state}
\label{subsec:merged_results}

For the merged equation of state, we choose the critical temperature
$T_c = 126.1~\mathrm{MeV}$ which leads to a critical baryon chemical potential
$\mu_B^c = 598~\mathrm{MeV}$, consistent with the critical chemical potential
predicted by the EMD model~\cite{Hippert:2023bel}.
The phenomenological merger parameter is set to
$\Delta V = 1.98\times10^{-6}~\mathrm{MeV}^{-3}$, which produces a smooth
transition between the input descriptions in the crossover
region~\cite{Yang:2026brr}. The equilibrium mixing weight $\bar p$ denotes the fraction associated with
the EMD equation of state. At low baryon chemical potential, $\bar p$ evolves
continuously from values close to zero at low temperature to values close to
unity at high temperature. The merged equation of state therefore changes
smoothly from the QvdW-HRG description of hadronic matter to the EMD
description of the deconfined phase. For
$\mu_B>\mu_B^c$ and $T<T_c$, the equilibrium mixing weight becomes
discontinuous, signaling a first-order transition between the two
descriptions.

The merged pressure closely follows the input equation of state with the
larger pressure in its corresponding regime of applicability.
\begin{figure}[h!]
\begin{subfigure}
     \centering
     \includegraphics[width=0.45\linewidth]{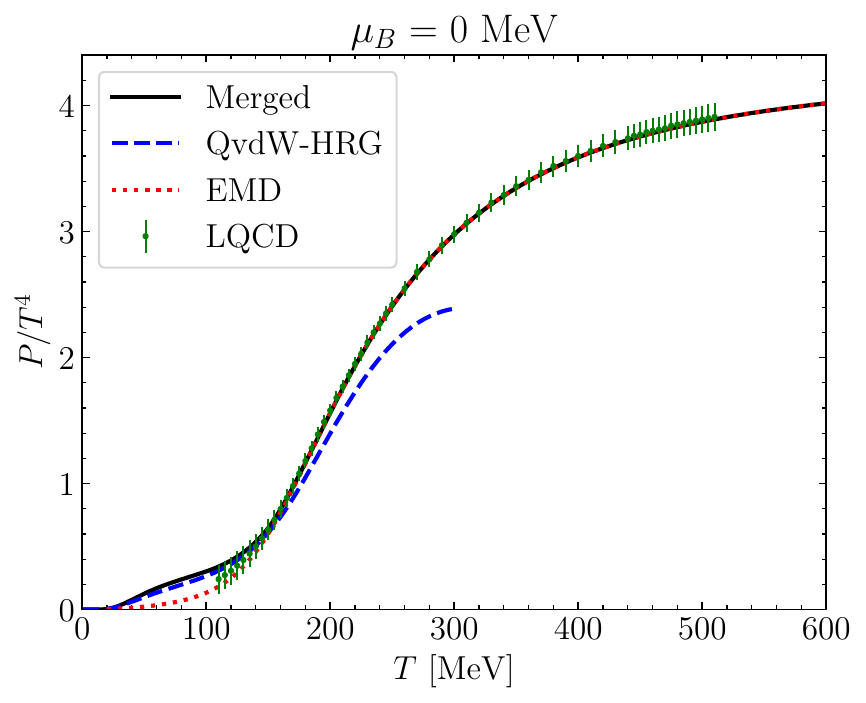}
\end{subfigure}
\begin{subfigure}
     \centering
     \includegraphics[width=0.45\linewidth]{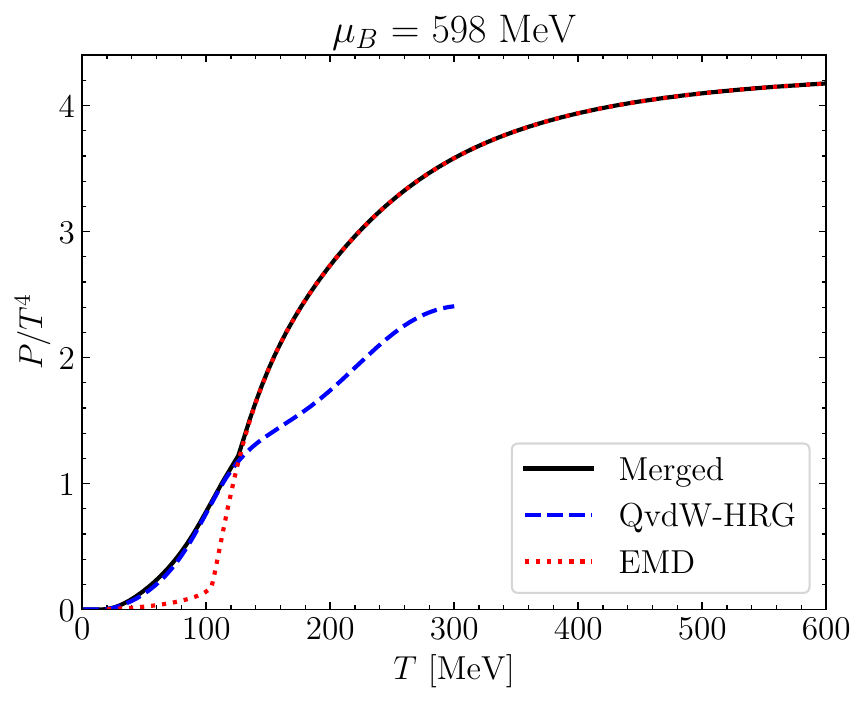}
\end{subfigure}
\caption{Pressure as a function of temperature at $\mu_B=0$ and
$\mu_B=598~\mathrm{MeV}$. At $\mu_B=0$, the results are compared with
lattice-QCD data from Ref.~\cite{Borsanyi:2013bia} (green points). In both
panels, the solid black line denotes the merged equation of state, while the
blue dashed and red dotted lines denote the QvdW-HRG and EMD equations of
state, respectively. Adapted from Ref.~\cite{Yang:2026brr}.}
    \label{fig:pressure}
\end{figure}
Below the critical chemical potential, the pressure, entropy density, baryon
density, and energy density interpolate continuously between the two models.
Above $\mu_B^c$, the pressure develops a kink, while the extensive
thermodynamic quantities become discontinuous across the first-order
transition. At the critical point, fluctuations of the mixing weight generate
the expected divergence of the second-order baryon susceptibility and the
vanishing of the speed of sound squared.

At $\mu_B=0$, the merged entropy and energy densities agree well with
continuum-extrapolated lattice-QCD results~\cite{Borsanyi:2013bia}.
Comparisons with finite-density lattice-QCD results obtained using the $T'$
expansion also show good agreement for the pressure, entropy density, baryon
density, and energy density~\cite{Borsanyi:2021sxv}. The result is a single
thermodynamically consistent equation of state spanning the hadronic and
deconfined regimes up to approximately $\mu_B=1~\mathrm{GeV}$ and
$T=600~\mathrm{MeV}$.
\section{Conclusions and outlook}
\label{sec:conclusions}
We presented a thermodynamically consistent and stable framework for merging
independent equations of state into a single global description. The method
introduces an internal order-parameter-like variable determined by minimizing
a single grand-potential density. It can describe a smooth crossover, a critical point with mean-field Ising behavior, and a first-order transition. As a proof of principle, we merged QvdW-HRG and holographic EMD equations of
state, obtaining a unified description of the hadronic and deconfined regimes
up to approximately $T=600~\mathrm{MeV}$ and
$\mu_B=1~\mathrm{GeV}$. The merged equation of state agrees well with
available lattice-QCD results and reproduces the characteristic thermodynamic
behavior near the critical point and first-order transition. Future extensions
include alternative input equations of state, electric-charge and strangeness
chemical potentials, and applications to dynamical simulations of baryon-rich
heavy-ion collisions.
\section{Acknowledgments}{
We thank V.~Vovchenko and H.~Shah for their help obtaining an equation of state for the hadron-resonance gas with the features  required for this work.  
This material is based upon work supported by
the National Science Foundation under grants No. PHY-2208724, PHY-2116686, PHY-2514763, PHY-2621752 and PHY-2623480, and within the framework of the MUSES collaboration, under grant number No. OAC-2103680. This material is also based upon work supported by the U.S. Department of Energy, Office of Science, Office of Nuclear Physics, under Award Number DE-SC0022023 and by the National Aeronautics and Space Agency (NASA)  under Award Number 80NSSC24K0767.
M.H. was supported by the Brazilian National Council for Scientific and Technological Development (CNPq) under process No. 313638/2025-0. 
Y.Y. and J.N. are partly supported by
the U.S. Department of Energy, Office of Science,
Office for Nuclear Physics under Award No. DE-
SC0023861. R.R. acknowledges financial support by CNPq under grants number 407162/2023-2 and 305466/2024-0.}


\bibliographystyle{elsarticle-num}
\bibliography{sqm2026_template}


\end{document}